\documentclass{PoS}
\usepackage{natbib}
\usepackage{graphicx}

\title{{\em NuSTAR} Observations of X-Ray Binaries}

\ShortTitle{NuSTAR Observations of X-Ray Binaries}

\author{\speaker{John A. Tomsick}\thanks{This talk was given on behalf of 
the {\em NuSTAR} Galactic Binaries Working Group.}\\
        Space Sciences Laboratory, 7 Gauss Way, University of California, 
Berkeley, CA 94720-7450, USA\\
        E-mail: \email{jtomsick@ssl.berkeley.edu}}

\author{Eric Bellm, Felix Fuerst, Fiona Harrison, Hiromasa Miyasaka, Shriharsh Tendulkar\\
        Cahill Center for Astronomy and Astrophysics, Caltech, Pasadena, CA 91125, USA\\
        E-mail: \email{fiona@srl.caltech.edu}}

\author{Varun Bhalerao\\
        Inter University Center for Astronomy and Astrophysics, P.O. Bag 4, 
Ganeshkhind, Pune 411007, India\\
        E-mail: \email{varunb@iucaa.ernet.in}}

\author{Deepto Chakrabarty\\
        Kavli Institute for Astrophysics and Space Research, MIT, 70 Vassar St., 
Cambridge, MA 02139, USA\\
        E-mail: \email{deepto@mit.edu}}

\author{Ashley King\thanks{Einstein Fellow}\\
        Department of Physics, 382 Via Pueblo Mall, Stanford, CA 94305, USA\\
        E-mail: \email{ashking@stanford.edu}}

\author{Jon M. Miller\\
        Department of Astronomy, University of Michigan, 1085 S. University Ave., 
Ann Arbor, MI 48109-1107, USA\\
        E-mail: \email{jonmm@umich.edu}}

\author{Lorenzo Natalucci\\
        Istituto Nazionale di Astrofisica, INAF-IAPS, via del Fosso del Cavaliere, 
I-00133, Roma, Italy\\
        E-mail: \email{Lorenzo.Natalucci@iaps.inaf.it}}

\author{Daniel Stern\\
        Jet Propulsion Laboratory, Caltech, Pasadena, CA 91109, USA\\
        E-mail: \email{daniel.k.stern@jpl.nasa.gov}}

\abstract{As of 2014 August, the {\em Nuclear Spectroscopic Telescope Array
(NuSTAR)} had observed $\approx$30 X-ray binaries either as part of the 
planned program, as targets of opportunity, or for instrument calibration.  
The main science goals for the observations include probing the inner part 
of the accretion disk and constraining black hole spins via reflection 
components, providing the first observations of hard X-ray emission 
from quiescent Low Mass X-ray Binaries (LMXBs), measuring cyclotron lines 
from accreting pulsars, and studying type I X-ray bursts from neutron stars.  
Here, we describe the science objectives in more depth and give an overview
of the {\em NuSTAR} observations that have been carried out to achieve the
objectives.  These include observation of four ``IGR'' High Mass X-ray 
Binaries (HMXBs) discovered by {\em INTEGRAL}.  We also summarize the results 
that have been obtained and their implications.  Among the IGR HMXBs, we focus 
on the discovery of a cyclotron line in the spectrum of IGR~J17544--2619.}

\FullConference{10th INTEGRAL Workshop: "A Synergistic View of the High Energy Sky" - Integral2014,\\
		15-19 September 2014\\
		Annapolis, MD, USA }

\begin{document}

\section{Introduction}
The {\em NuSTAR} satellite launched in 2012 June, and has been carrying out
science operations for almost 2 years (at the time of the conference).  
{\em NuSTAR} observes in the 3--79\,keV bandpass and is the first focusing 
hard X-ray telescope in orbit.  Its sensitivity, angular resolution 
($58^{\prime\prime}$ Half-Power Diameter), and energy resolution above 10\,keV 
are unprecedented \citep{harrison13}.  {\em NuSTAR}'s field of view is 
12$\times$12\,arcmin$^{2}$, which precludes an all-sky survey.  Thus, 
the standard {\em NuSTAR} operation is to point from source-to-source, and 
a detailed observing plan is required.

The mission has broad science goals across many areas of astrophysics.  
The good angular and energy resolution in the hard X-ray band (900\,eV FWHM 
at 60\,keV) allows for a sensitive study of $^{44}$Ti emission lines from 
supernova remnants like Cas A \citep{grefenstette14}.  The sensitivity, 
angular resolution, and timing capabilities have also been used to find
and identify stellar remnants near the Galactic center \citep{mori13}
and in the Galactic plane \citep{gotthelf14}.  {\em NuSTAR}'s sensitivity 
is extending our reach into the Universe by detecting fainter Active 
Galactic Nuclei \citep{alexander13}.

Here we report on the {\em NuSTAR} program for observing X-ray binaries
in the Galaxy.  Section 2 introduces the topics and science objectives
that we used in guiding our observing program.  Section 3 summarizes
the observations that had been carried out as of 2014 August.  Section 4
summarizes results to date, including follow-up observations of the
intermediate luminosity sources discovered by {\em INTEGRAL}.

\section{Science Objectives for X-ray Binaries}
In designing the X-ray binaries program, the Galactic Binaries Working
Group has considered the full array of {\em NuSTAR} capabilities.  In 
considering the most important advances that {\em NuSTAR} provides, we
divided our program into the following four topics:

{\bf 1. Reflection from black hole accretion disks:}  A Compton reflection
component is seen when a hard X-ray source illuminates the accretion
disk, and this produces fluorescent iron line emission, iron absorption
edges, and a hard X-ray reflection hump \citep{lw88,fabian89}.  These 
features are distorted by Doppler and gravitational effects when the disk 
approaches the black hole, and allows for a measurement of the geometry of 
the source and the accretion disk.  {\em NuSTAR}'s ability to do this 
measurement was demonstrated early-on in the mission for AGN 
\citep{risaliti13}.  Once it was clear that the {\em NuSTAR} bandpass 
would extend down to 3\,keV, we saw that this would allow for a measurement 
of this full reflection component, and several observations of black holes
and neutron stars have been carried out with a reflection measurement as
a goal (see below).  The targets have been mostly bright sources (e.g., 
Cyg~X-1, GRS~1915+105, Ser~X-1), and the throughput of {\em NuSTAR}'s
CZT detectors has been of critical importance.

{\bf 2. Quiescent Low Mass X-ray Binaries:}  While X-ray binaries are bright
when they are in outburst, most accreting black holes and many accreting 
neutron stars are transient, and become very faint when they are in 
quiescence.  With {\em Chandra} and {\em XMM-Newton}, it became possible 
to detect the majority of the quiescent transients and to study several in 
detail in the soft X-ray band, no previous satellite could observe them
in the hard X-rays.  With {\em NuSTAR}'s sensitivity, we have been able
to observe some of these systems for the first time.

{\bf 3. High Mass X-ray Binaries and accreting pulsars:}  For most of the 
known High Mass X-ray Binaries (HMXBs), the accreting compact object is
a neutron star with a relatively high magnetic field.  These systems
often exhibit pulsations, they sometimes show Cyclotron Resonant
Scattering Features or ``cyclotron lines'' \citep{tru78,coburn02}, and
they typically have a continuum that is very hard, peaking near 15--25\,keV
with a high-energy cutoff.  The cyclotron lines are dips in the energy
spectrum at energies which are directly and linearly related to the neutron 
star magnetic field.  A magnetic field strength of $10^{12}$\,G corresponds
to a line at 12\,keV, placing line energies for most HMXBs in the middle
of the {\em NuSTAR} bandpass.  With {\em NuSTAR}'s energy resolution and
sensitivity, more detailed studies of cyclotron lines are possible, and 
they are detectable for more of the fainter systems.  As {\em INTEGRAL}
has led to a large increase in the HMXB population, and many of the
``IGR'' HMXBs have lower fluxes, an important part of our program is
to study IGR HMXBs.

{\bf 4. Type I X-ray bursts:}  Thermonuclear X-ray flashes occur in systems 
where accretion onto a low magnetic field ($\sim$$10^{8-9}$\,G) neutron star
occurs.  Typically, the burst spectrum is a $\sim$1--3\,keV blackbody where
the temperature evolves while the flux changes by orders of magnitude over
a time period of $\sim$10--100\,s \citep{lewin93}.  Some systems show bursts
that are powerful enough to cause expansion of the neutron star's photosphere, 
and a fraction of these are ``superexpansion'' bursts where the photosphere
increases its radius by a factor of $\sim$100.  It is in these superexpansion
bursts where absorption edges may be present.  With the energy resolution
of the {\em Rossi X-ray Timing Explorer} Proportional Counter Array, these
features appear as maxima in the residuals near 6--7\,keV and minima near
10--13\,keV \citep{iw10}, and there are heavy elements (Fe, Co, Zn, Ni), 
which have absorption edges that could cause the observed features.  With
the improvement in energy resolution, {\em NuSTAR} will be able to 
confirm the nature of these features and measure the edge energies.  

\section{Overview of {\em NuSTAR} Observations}
\begin{figure}
\centerline{\includegraphics[width=5.5in,angle=0]{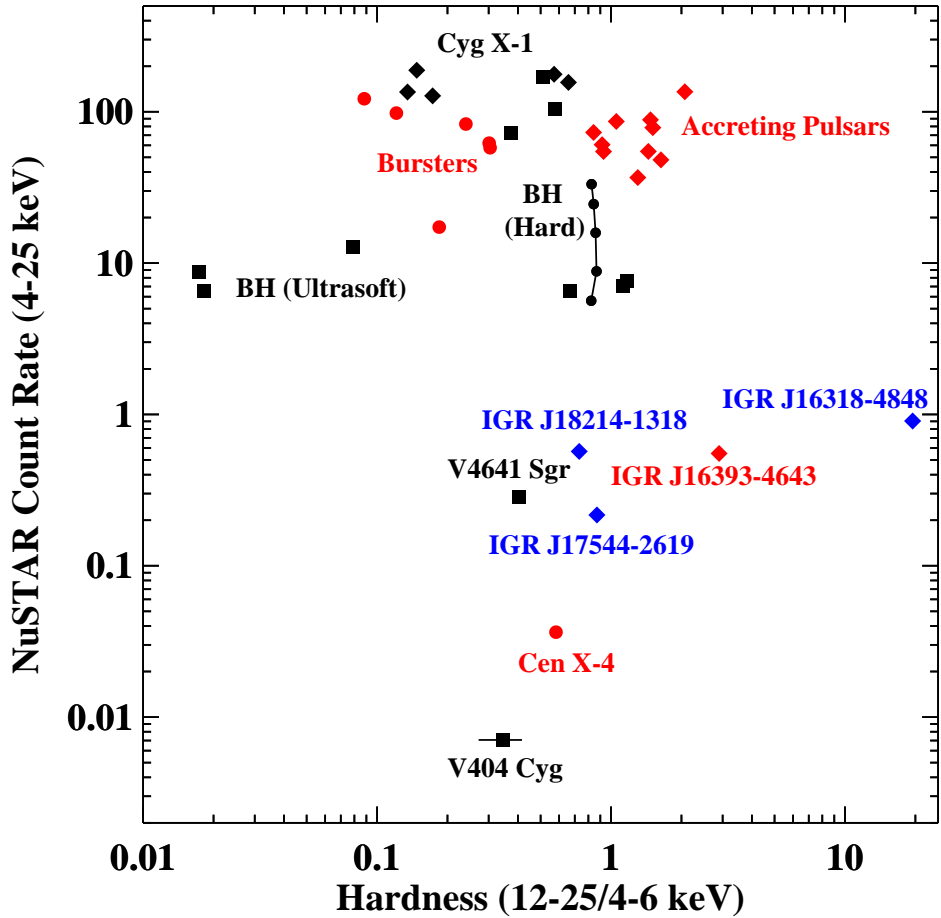}}
\caption{Hardness-intensity diagram for X-ray binaries (black points
are accreting black holes and red points are accreting neutron stars)
observed by {\em NuSTAR} through 2014 August. The black diamonds are
observations of Cyg~X-1 in the hard and soft state, and the five 
connected circles are observations of GX~339--4 in the hard state.
\label{fig:colors}}
\end{figure}

For most of the {\em NuSTAR} observations carried out through 2014 August, 
we have extracted count rates in the 4--6\,keV, 6--12\,keV, and 12--25\,keV
energy bands, and Figure~\ref{fig:colors} shows a hardness-intensity 
diagram where each point represents a individual {\em NuSTAR} observation.  
There are rates from 33 X-ray binaries shown on the plot, and most of them
have been observed one or two times.  The sources that have been most frequently
observed are the accreting black holes Cyg~X-1 and GX~339--4.  The exposure 
times are typically $\sim$30\,ks per observation, but they range from a few ks 
(especially for some of the observations during in-orbit checkout) to $\sim$100\,ks
for the quiescent LMXBs.

Figure~\ref{fig:colors} gives some indication of how much time we have spent
on each of our science objectives outlines above.  Most of the black hole 
observations are related to Goal \#1 (reflection).  For Goal \#2 (quiescent
LMXBs), we have observed Cen~X-4 and V404~Cyg, and these are the brightest
of the quiescent LMXBs.  We have observed several accreting pulsars for 
Goal \#3 (cyclotron lines), and some bursters for Goal \#4 (absorption edges).

The observations take advantage of the very large dynamic range of {\em NuSTAR}, 
which can make useful observations of sources with count rates extending over 
four orders of magnitude.  The limit at the high count rate end is caused by
deadtime, but, even for a source as bright as Cyg~X-1, the {\em NuSTAR} deadtime
is much lower than a CCD instrument \citep{tomsick14}, there is no significant
pile-up, and the spectrum is not distorted.  The low count rate limit depends
on the science goals.  Our goals at least involved measuring the energy spectrum 
and obtaining a detection above 10\,keV, and for this, the V404~Cyg rate is
close to the limit for a $\sim$100\,ks observation.  The location of the 
four IGR HMXBs that we have observed are at intermediate count rates on the 
hard side of the plot.  This is a region of parameter space where {\em NuSTAR}
exposures of moderate duration can make a huge improvement over previous
measurements, and we provide a more detailed example below.

\section{Results and Discussion}
Here, we provide a brief status report on the results that have been obtained 
for each of the four goals listed above.  Several papers are in preparation, 
but we restrict ourselves to describing published work. 

\subsection{Goal \#1:  Reflection}
There are three black hole systems for which we have used modeling of the reflection
component to constrain the spin of the black hole:  Cyg~X-1, GRS~1915+105, and 
4U~1630--47.  For Cyg~X-1, the study used 30\,ks {\em NuSTAR} and {\em Suzaku} 
observations from 2012 October 31 - November 1, covering the 1--300\,keV bandpass 
\citep{tomsick14}.  The source was in the soft state, and the spectrum includes a 
multi-temperature blackbody, a power-law, and a strong reflection component.  The
Fe K$\alpha$ emission line is very broad and the line and reflection component are
consistent with the relativistic smearing expected if the emission is coming from 
the inner accretion disk.  Assuming that the disk is at the innermost stable 
circular orbit (ISCO), we obtained a black hole spin of $a_{*} > 0.83$ \citep{tomsick14}, 
which is consistent with previous measurements using the reflection component
\citep{duro11,miller12,fabian12} and using a method of modeling the thermal
continuum \citep{gou14}.  The reflection modeling also provides a constraint on
the inclination of the inner disk, and we find that $i > 40^{\circ}$ \citep{tomsick14}, 
which is significantly higher than the value of $i = 27.1^{\circ}\pm 0.8^{\circ}$ 
obtained for the binary inclination \citep{orosz11}.  This may indicate that there 
is a warp in the accretion disk.

We also used reflection modeling to measure the black hole spins for GRS~1915+105
and 4U~1630--47.  For GRS~1915+105, the {\em NuSTAR} measurement with 1-$\sigma$
statistical errors is $a_{*} = 0.98\pm 0.01$ \citep{miller13}, and this compares well
with the previous value found with thermal modeling \citep{mcclintock06}.  For
4U~1630--47, the value (also with 1-$\sigma$ statistical errors) is 
$a_{*} = 0.985^{+0.005}_{-0.014}$ \citep{king14}, and this is the first time a black
hole spin measurement has been obtained for this system.  In addition to the broad
emission line, absorption was also seen in the system, and \cite{king14} interpret
this as a blue-shifted absorption line from a high-velocity disk wind.  

We obtained {\em NuSTAR} observations of 1E~1740.7--2942, which is an accreting 
black hole and ``microquasar'' (an X-ray binary with radio jets) within 
$\approx$$1^{\circ}$ of the Galactic Center.  The source was in the hard state when 
we observed it, which is its most common state.  As the source is in a very 
crowded region, {\em NuSTAR} provides an advance since we are able to definitively
obtain an uncontaminated hard X-ray spectrum.  We measured a spectrum that is
well-described by thermal Comptonization, and we did not detect a reflection
component with 90\% confidence upper limits on the covering fraction of 1--8\%, 
depending on the disk ionization \citep{natalucci14}.

In addition to the black hole results, we measured the reflection 
component for the neutron star LMXB Ser~X-1 with {\em NuSTAR}, marking the first
time that the hard X-ray reflection hump was detected in a neutron star system
\citep{miller13_serx1}.  As the inner disk edge is close to the neutron star 
surface, such measurements may be promising for obtaining a constraint on the
neutron star radius.

\subsection{Goal \#2:  Quiescent LMXBs}
{\em NuSTAR} observed the neutron star LMXB Cen~X-4 for 114\,ks in 2013 January.
Part of the observation was also covered by {\em XMM-Newton}, and the results
were reported in \cite{chakrabarty14}.  Previous observations with {\em XMM-Newton}
and other soft X-ray satellites of Cen~X-4 in quiescence show that the spectrum has 
thermal emission with a temperature of $\sim$0.1\,keV from the neutron star surface
as well as a hard component that is consistent with being a power-law with an
index of $\Gamma = 1$--2 \citep{cackett10}.  The power-law component has never been 
detected above 10\,keV, and a major question has been where the cutoff to the
component occurs.  The {\em NuSTAR} observation shows a spectrum that can be fitted
with an exponential cutoff at $10.4\pm 1.4$\,keV (1-$\sigma$ uncertainties), and we 
found that the hard component can also be explained as $kT = 18.2\pm 1.0$\,keV
bremsstrahlung emission \citep{chakrabarty14}.  We considered the physical implications
of the hard component originating from inverse Compton, synchrotron shock, and 
bremsstrahlung mechanisms and found that inverse Compton and synchrotron shock
are ruled out on physical grounds.  We favor a scenario where the hard component
is bremsstrahlung emission from a hot layer above the neutron star atmosphere
\citep{chakrabarty14,dangelo14}.

\subsection{Goal \#3:  Accreting Pulsars and Cyclotron Lines}
Thus far, {\em NuSTAR} measurements of cyclotron lines have been reported for the 
accreting pulsars Her~X-1 \citep{fuerst13}, Vela~X-1 \citep{fuerst14_velax1}, 
KS~1947+300 \citep{fuerst14_1947}, and RX~J0520.5--6932 \citep{tendulkar14}.  
While they all have relatively typical magnetic field strengths of a few times
$10^{12}$\,G, {\em Suzaku} reported that GRO~J1008--57 has a cyclotron line near
80\,keV \citep{yamamoto14}, and a combined {\em NuSTAR} and {\em Suzaku} study 
confirmed the presence of a line at $78^{+3}_{-2}$\,keV \citep{bellm14}.  We also 
used the {\em NuSTAR} spectrum to look for the possibility of a fundamental
cyclotron line at lower energies (e.g., if the 78\,keV line is a harmonic, then
we might find a fundamental near 39\,keV).  However, we did not find any evidence
for a line at lower energy, providing strong evidence that the 78\,keV line is
the fundamental, indicating a neutron star magnetic field strength near 
$7\times 10^{12}$\,G \citep{bellm14}.

\begin{figure}
\centerline{\includegraphics[width=4.5in,angle=0]{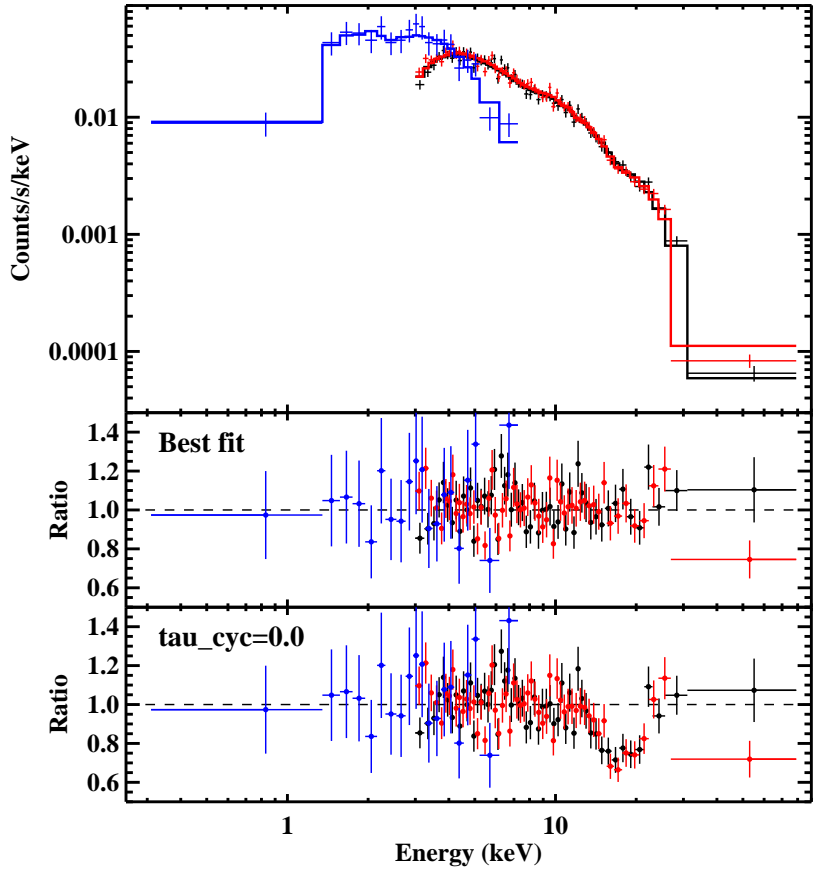}}
\caption{{\em Swift}/XRT and {\em NuSTAR} spectrum of the Supergiant
Fast X-ray Transient IGR~J17544--2619.  The observations occurred on
2013 June 18-19, and the {\em NuSTAR} exposure time was 27\,ks
\citep{bhalerao14}.  The top panel shows the count spectrum fitted 
with a blackbody, a cutoff power-law, and a cyclotron line.  The
middle panel shows the residuals to that model, and the bottom
panel shows the residuals when the cyclotron line optical depth
is set to zero.\label{fig:j17544}}
\end{figure}

A {\em NuSTAR} observation of the HMXB IGR~J17544--2619 led to the first detection
of a cyclotron line in a Supergiant Fast X-ray Transient (SFXT).  SFXTs have 
unusually bright flares, and it has been suspected that these might be related
to very strong neutron star magnetic field strengths \citep{bfs08}.  However, 
the line energy is $16.8\pm 0.3$\,keV \citep{bhalerao14}, which corresponds to a 
typical value for a neutron star HMXB.  Figure~\ref{fig:j17544} shows the 
{\em NuSTAR} and {\em Swift}/XRT spectra from an observation that occurred on 
2013 June 18-19 fitted with a $\sim$1\,keV blackbody, a power-law with an 
exponential cutoff, and a cyclotron line.  More than one continuum model can 
provide a good fit to the spectrum, but the cyclotron line is always required.  
Simulations were performed to determine that 
the statistical significance of the line is $>$5-$\sigma$.  The spin period
of the IGR~J17544--2619 neutron star is still unclear, but knowing this would
be very helpful in understanding the production of the large flares from SFXTs.

Although the {\em NuSTAR} observation of GS~0834--430 did not lead to a cyclotron
line detection, we studied the energy dependence of the 12.3\,s pulsations from
this accreting pulsar.  The pulse profiles show a very strong hard phase lag
with a shift by 0.3 cycles going across the {\em NuSTAR} bandpass \citep{miyasaka13}.  
This was the first time that such large energy-dependent lags have been reported
for an HMXB, and \cite{miyasaka13} conclude that the most likely explanation for
the lags is a complex pulsar beam geometry.

\subsection{Goal \#4:  Type I X-ray Bursts}
For our observing program, we selected bursters that are known to enter states where 
bursts occur frequently (e.g., every few hours) to provide a good chance that
{\em NuSTAR} will detect bursts.  Furthermore, with detection of absorption edges 
during superexpansion bursts being our primary goal, we selected Ultracompact
X-ray Binaries (UCXBs) since they are known to be more likely to produce
such bursts.  We planned a Target of Opportunity for when either 4U~1820--30
or 4U~1728--34 entered into a state where superexpansion bursts are likely to
occur.  4U~1820--30 was observed briefly on 2013 July 8, but it was not in the
correct state.  4U~1728--34 was observed on 2013 October 1 for 33\,ks and 
on 2013 October 3 for 33\,ks, and bursts were detected, but they were not
superexpansion bursts.  For a {\em NuSTAR} paper on a serendipitous detection
of a type I X-ray burst from GRS~1741.9--2853, we refer the reader to 
\cite{barriere14}.

\section{Conclusions}
{\em NuSTAR}'s combination of sensitivity, throughput, and energy resolution is 
providing significant new results in studies of X-ray binaries.  The large 
dynamic range that {\em NuSTAR} provides allows for useful observations of bright 
(e.g., Cyg~X-1) and faint (e.g., Cen~X-4) systems.  Some of the largest advances 
over previous satellites can be made with moderate-length observations of sources 
with intermediate count rates, and IGR HMXBs provide excellent examples.  This is
illustrated especially by the discovery of a cyclotron line in the SFXT
IGR~J17544--2619.


\end{document}